\documentclass[aps,pre,twocolumn,twoside, amsmath,amssymb,,superscriptaddress,
 aps,bbm,floatfix,showpacs]{revtex4-1}

\usepackage{xcolor}
\usepackage{graphicx}
\usepackage{dcolumn}
\usepackage{bm}
\usepackage{framed}
\usepackage{hyperref}
\usepackage{dsfont}
\usepackage[normalem]{ulem}
\usepackage{nicefrac}

\begin{document}

\title{Quantum witness of a damped qubit with  generalized measurements}
\author{Manuel Bojer}
\author{Alexander Friedenberger} 
\affiliation{Department of Physics, Friedrich-Alexander-Universit\"at Erlangen-N\"urnberg, D-91058 Erlangen, Germany}

\author{Eric Lutz}
\affiliation{Department of Physics, Friedrich-Alexander-Universit\"at Erlangen-N\"urnberg, D-91058 Erlangen, Germany}

\affiliation{Institute for Theoretical Physics I, University of Stuttgart, D-70550 Stuttgart, Germany}

\begin{abstract}
We evaluate the quantum witness based on the no-signaling-in-time condition of a damped two-level system for nonselective generalized measurements of varying strength. We explicitly compute its dependence  on the measurement strength for a generic example. We find a vanishing derivative for  weak measurements and an infinite derivative in the limit of projective measurements. The quantum witness is hence mostly insensitive to the strength of the measurement in the weak measurement regime and displays a singular, extremely sensitive dependence for strong measurements. We finally relate this behavior  to that of the measurement disturbance defined in terms of the fidelity  between pre-measurement and post-measurement states.

\end{abstract}

\maketitle

\section{Introduction}
\label{sec:Introduction}
What is the intrinsic difference between quantum and classical systems? This question has been at the core of quantum theory since its very beginning \cite{leg02,rei09}. From a practical point of view, nonclassicality has  been recognized as a useful resource for quantum technologies  that can  outperform their classical counterparts, including quantum communication \cite{gis07}, quantum computation \cite{vin95} and quantum metrology \cite{gio11}. Quantum properties have been assessed in the past in the form of violations of inequalities due to Bell \cite{bel64,bru14} and Leggett and Garg \cite{leg85,ema14}. A violation of  the former  discloses the presence of  nonclassical spatial correlations between two systems, whereas a  violation of the latter reveals nonclassical temporal correlations in the evolution of a single system. Recently, a third criterion based on the no-signaling-in-time condition  has been developed \cite{nor12,kof12}. This approach exploits the idea that the time evolution of a quantum system is perturbed by a measurement contrary to that of a classical system. In this context, quantum features are  witnessed by the different population dynamics  of a system in the presence and  in the absence of a measurement. The advantage of this quantity is its simplicity, since one has only to prepare the initial state of the system and perform two different measurements.  Experimental investigations of such quantum witness with single atoms \cite{rob15},   superconducting flux qubits \cite{kne16} and individual photons \cite{wan17,wan17a} have been reported lately. 

The above nonclassicality criteria are commonly studied with ideal (projective) measurements. However, general measurements on quantum systems are described by (non-projective) positive operators, $E_i\geq0$, that satisfy $\sum_i E_i = I$ with $I$ the identity \cite{jac14}. These operators are known as POVMs (positive-operator valued measures) and are not orthogonal contrary to projectors. Most actual experiments do not correspond to projective measurements, but to some more general POVMs. The influence of POVMs on Bell nonlocality has been investigated both theoretically \cite{ver10,bar12,hir13,qui15,ozm17,hir18} and experimentally \cite{gom16,gom18}. At the same time, theoretical \cite{rus06,mal16,kum17} and experimental \cite{pal10,whi15} studies of violations of the Leggett-Garg inequality with generalized measurements have been performed.

Our aim in this paper is to investigate the quantum witness based on the no-signaling-in-time condition of a damped qubit when a generalized measurement with variable strength is used to perturb the system. We explicitly compute the quantum witness for a  two-level system with eigenbasis $\sigma_z$ and   an incomplete nonselective measurement in the $\sigma_x$-basis by analytically solving the corresponding Lindblad master equation. We concretely determine the dependence of the witness on the measurement strength. We obtain a vanishing derivative for weak measurements that indicates that the quantum witness is insensitive to the measurement strength in that limit. We further find a divergent, non-analytic derivative for strong measurements which reveals a singular, highly sensitive dependence  on the strength of the measurement. We finally relate this behavior to that of the measurement disturbance defined in terms of the fidelity 
between the states before and after the measurement.

\section{Witness of a damped qubit}
\label{sec:Two_lev} 
We consider a two-level system  with Hamiltonian $H = {\omega}\sigma_z/2$, where $\omega$ is the transition frequency  and $\sigma_z$ the $z$-Pauli operator, weakly  coupled to  a thermal radiation field at temperature $T$ ($\hbar = 1$ throughout for simplicity).  In the Born-Markov approximation, the time evolution in the Heisenberg picture of a system observable $X(t)$ obeys a Lindblad master equation of the form \cite{bre02},
\begin{equation}
\begin{aligned}
\frac{dX}{dt}&=\mathcal{L}[X]=i\frac{\omega}{2} \left[\sigma_z,X\right]+\frac{\partial X}{\partial t} \\
&+{\frac{\gamma_{0}}{2}n(\omega,T)}\left(\sigma_{-}\left[X,\sigma_{+}\right] +\left[\sigma_{-},X\right]\sigma_{+}\right)\\
&+{\frac{\gamma_{0}}{2}(n(\omega,T)+1)}\left(\sigma_{+}\left[X,\sigma_{-}\right] +\left[\sigma_{+},X\right]\sigma_{-}\right),\label{eq:mastereq}
\end{aligned}
\end{equation}
where $\sigma_{\pm}$ are the raising and lowering operators of the qubit, $\gamma_0$  the spontaneous decay rate and $n(\omega,T)= [\exp(\omega/k_B T)-1]^{-1}$ the thermal occupation number. We denote the total transition rate by $\gamma=\gamma_0\left[2 n(\omega,T)+1\right]$.

The master equation \eqref{eq:mastereq} may be solved by using a superoperator formalism \cite{Crawford1958,muk95}.  By choosing a basis consisting of the Pauli operators $\sigma_i\, (i=x ,y, z)$ and the identity operator $I$,  the matrix representation of the Liouvillian superoperator  $\mathcal{L}$ is,
\begin{equation}
\frac{d}{dt}\begin{pmatrix}
 \sigma_x\\
\sigma_y\\
\sigma_z\\
I
\end{pmatrix}= \mathcal{L}\begin{pmatrix}
\sigma_x\\
\sigma_y\\
\sigma_z\\
I
\end{pmatrix}= 
{\begin{pmatrix}
-\frac{\gamma}{2}&-\omega&0&0\\
\omega&-\frac{\gamma}{2}&0&0\\
0&0&-\gamma&-\gamma_0\\
0&0&0&0
\end{pmatrix}}\begin{pmatrix}
\sigma_x\\
\sigma_y\\
\sigma_z\\
I
\end{pmatrix}.\label{2}
\end{equation}
The formal solution of Eq.~\eqref{2} can be written after time integration in terms of the propagator,
\begin{equation}
\label{3}
e^{\mathcal{L}t} = \left(\begin{array}{cccc} {e}^{-\frac{\gamma t}{2}}\,\cos\omega t & -{e}^{-\frac{\gamma t}{2}}\,\sin\omega t & 0 & 0\\ {e}^{-\frac{\gamma t}{2}}\,\sin\omega t & {e}^{-\frac{\gamma t}{2}}\,\cos\omega t & 0 & 0\\ 0 & 0 & {e}^{-\gamma t} & \frac{\gamma_0\left({e}^{-\gamma t}-1\right)}{\gamma}\\ 0 & 0 & 0 & 1 \end{array}\right)\,.
\end{equation}
The density operator $\rho(t)$ of the qubit then follows as,
\begin{equation}
\rho(t) = \frac{1}{2}\left(I + \left\langle \sigma_x (t)\right\rangle \sigma_x + \left\langle \sigma_y (t)\right\rangle \sigma_y + \left\langle \sigma_z (t)\right\rangle \sigma_z\right),
\end{equation} 
where $\langle \sigma_i(t)\rangle$ denotes the expectation value at time $t$ of the operator $ \sigma_i$ evaluated with Eq.~\eqref{3}.

In order to assess the nonclassicality of the two-level system, we next introduce the quantum witness based on the   no-signaling-in-time condition \cite{nor12,kof12}. Consider two system observables $A$ and $B$ that are respectively measured at $t=t_0$ and $t>t_0$. The measurement outcome $a_k$ of $A$ is obtained with probability $P(a_k)$ and the measurement outcome $b$ of $B$ with probability $P(b)$. For  a joint measurement of the two observables, the probability of finding the outcome $b$ in the second measurement is,
\begin{equation}
P'(b) = \sum_{k=1}^{M}{P(b,t\vert a_k,t_0)}P(a_k),
\end{equation}
where $P(b,t\vert a_k,t_0)$ describes the conditional probability of obtaining the result $b$ at time $t$ given the result $a_k$ at time $t_0$. We have   here assumed that there are $M$ possible measurement outcomes of the nonselective measurement of $A$. In the absence of the first measurement on $A$, we denote the probability of outcome $b$ of $B$  by  $P(b)$. The quantum witness is then defined as,
\begin{equation}
\label{6}
\mathcal{W}_q = \left|P(b)-P'(b)\right|\,.
\end{equation}
According to the classical no-signaling in time condition, the measurement done on $A$ should have no influence on the outcomes of the later measurement of $B$ \cite{nor12,kof12}. The quantum witness therefore vanishes for a classical qubit. A nonzero value, $\mathcal{W}_q > 0$, is a clear  signature of the nonclassicality of the system.

\begin{figure}[t]
\centering
\includegraphics[width=0.45\textwidth]{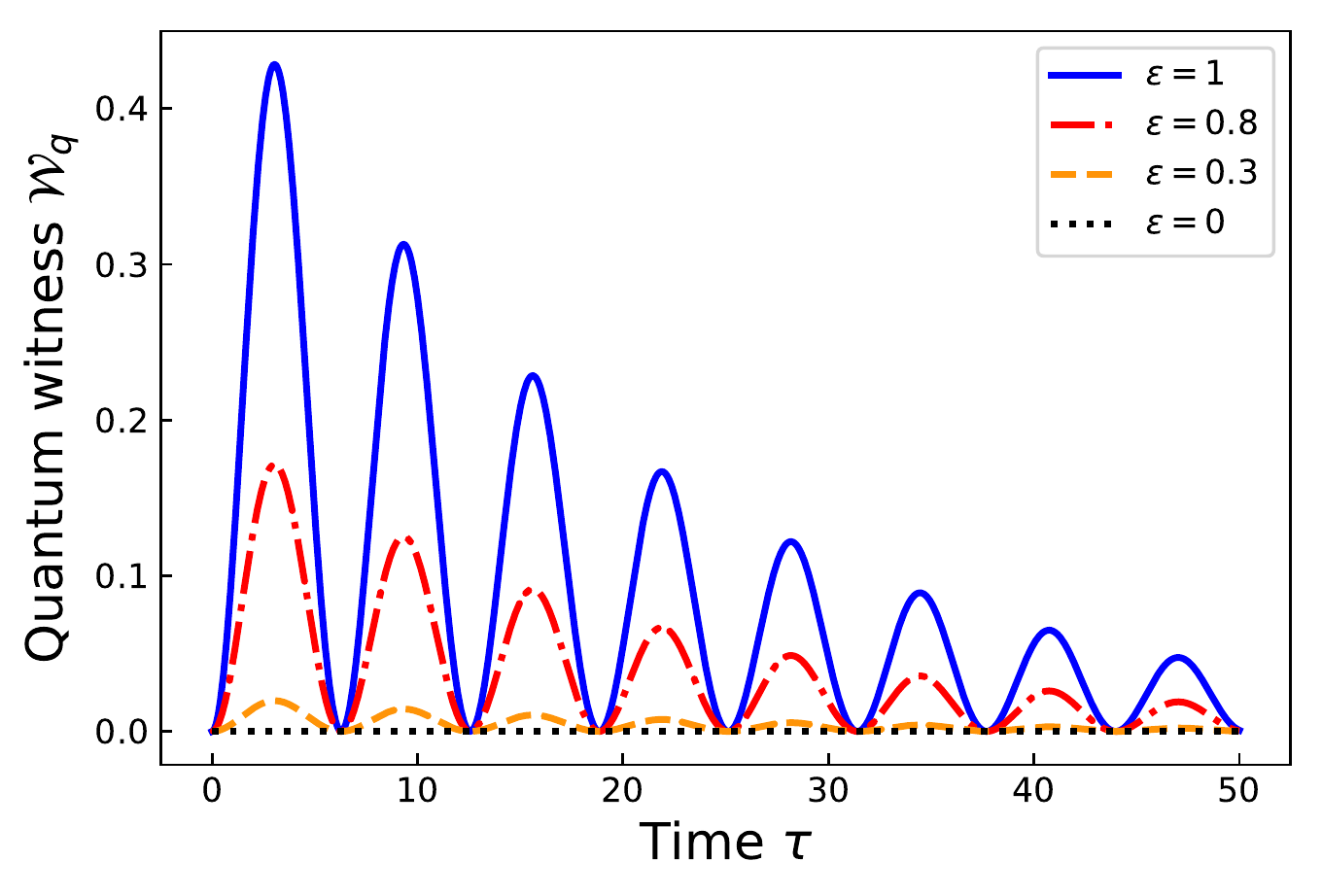}
\caption{Quantum witness $\mathcal{W}_q(\tau,\varepsilon)$, Eq.~\eqref{eq:Wit_weak}, of a damped two-level system as a function of   time $\tau$ for various measurement strengths $\varepsilon$ of the generalized measurement defined by the  positive operators (7)-(8). The oscillation amplitude of the witness is suppressed as a result of the combined effect of the external damping and of the finite measurement strength. Parameters are $\omega= 1$ and $\gamma= 0.1$.}
\label{fig:Weak_Measurement_Coupling}
\end{figure}

We evaluate the quantum witness \eqref{6}  for a damped two-level system with $M=2$ in the following way.
We assume that the system is initially prepared in the eigenstate  $|+\rangle =\left(|\uparrow\rangle +|\downarrow\rangle\right) /\sqrt{2} $ of the  operator $\sigma_x$ at $t = 0$ \cite{sch15,fri17}. At time $t= \tau/2$  a nonselective generalized measurement in the $\sigma_x$-basis is performed, or not, whereas at $t=\tau$ the  projector $\Pi_+ = \left|+\right\rangle\left\langle + \right|$ is measured. 
We characterize the generalized measurement with the help of the two positive operators \cite{bru02,doh00,fuc01},
\begin{align}
E_+ &= \left(\frac{1+\varepsilon}{2}\right) \left|+\right\rangle\left\langle +\right| + \left(\frac{1-\varepsilon}{2}\right) \left|-\right\rangle\left\langle -\right|= A_+^2,\\
E_- &= \left(\frac{1-\varepsilon}{2}\right) \left|+\right\rangle\left\langle +\right| + \left(\frac{1+\varepsilon}{2}\right) \left|-\right\rangle\left\langle -\right|= A_-^2,
\end{align}
such that $E_+ + E_- = I$ and 
\begin{align}
A_+ &= \sqrt{\frac{1+\varepsilon}{2}} \left|+\right\rangle\left\langle +\right| + \sqrt{\frac{1-\varepsilon}{2}} \left|-\right\rangle\left\langle -\right|,\label{POVM1}\\
A_- &= \sqrt{\frac{1-\varepsilon}{2}} \left|+\right\rangle\left\langle +\right| + \sqrt{\frac{1+\varepsilon}{2}} \left|-\right\rangle\left\langle -\right|. \label{POVM2}
\end{align}
The parameter $\varepsilon$ ($0\leq \varepsilon \leq 1$) describes the finite strength of the measurement: $\varepsilon=1$ represents a (strong) projective measurement of the observable $\sigma_x$, whereas the limit $\varepsilon \ll 1$  corresponds to a weak measurement where the state of the system is slightly perturbed and only a small amount of information is gained by measuring it. When  $\varepsilon =0$, both operators (9) and (10) are proportional to the identity and  nothing is learned from the measurement. Incomplete measurement operators of this form appear quite naturally in the theory of continuous quantum measurements \cite{doh00}. They may also describe a finite-temperature Stern-Gerlach device that measures the spin of an electron in the presence of thermal noise  when $\varepsilon$ is replaced by $2\kappa -1$ ($-1/2\leq \kappa \leq 1/2$)\cite{fuc01}.
 
\begin{figure}[t]
\centering
\includegraphics[width=0.45\textwidth]{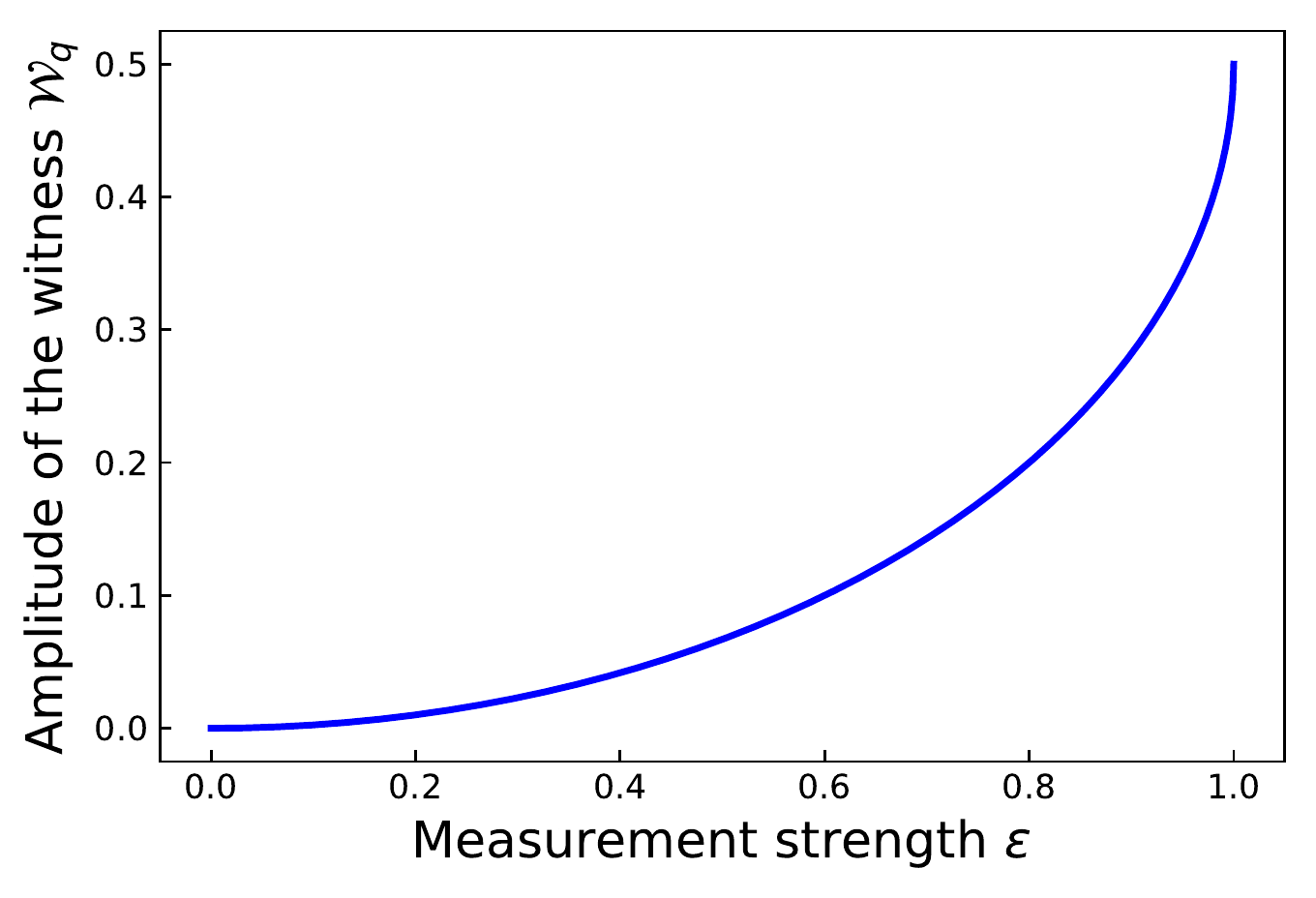}
\caption{Amplitude factor $f(\varepsilon) = 1-\sqrt{1-\varepsilon^2}$ of the quantum witness $\mathcal{W}_q(\tau,\varepsilon)$, Eq.~\eqref{eq:Wit_weak}, as a function of the measurement strength $\varepsilon$. A vanishing derivative is observed for weak measurements, $\varepsilon \rightarrow0$, while the derivative diverges for strong measurements, $\varepsilon \rightarrow1$. The latter property reveals a singular, highly sensitive, dependence on the measurement strength.}
\label{fig:Weak_Measurement_Coupling}
\end{figure}

The probability to find the system in state $|+\rangle$ at time $t=\tau$ in the absence of the intermediate nonselective generalized measurement is (see Appendix),
\begin{equation}
P(\left|+\right\rangle) = \left\langle +\right|\rho(\tau)\left|+\right\rangle = \frac{1}{2}\left(1+e^{-\frac{\gamma\tau}{2}}\cos\omega \tau\right),
\end{equation} 
while the corresponding probability in the presence of the nonselective generalized measurement reads,
\begin{equation}
P'\left(\left|+\right\rangle\right) = \frac{1}{2}\bigg[1 + \,e^{-\frac{\gamma\tau}{2}}\big(\cos^2\frac{\omega\tau}{2}  
- \sqrt{1-\varepsilon^2}\sin^2\frac{\omega\tau}{2}\big)\bigg].
\end{equation}
As a result, the  quantum witness \eqref{6} is given by,
\begin{equation}
\begin{aligned}
\mathcal{W}_q(\tau,\varepsilon)& = 
\frac{e^{-\frac{\gamma\tau}{2}}}{2} \left(1-\sqrt{1-\varepsilon ^2}\right)\sin^2\frac{\omega\tau}{2}.
\label{eq:Wit_weak}
\end{aligned}
\end{equation}
We observe that the quantum witness oscillates with frequency $\omega/2$ in the absence of damping and that the amplitude of these oscillations decays exponentially with decay constant $\gamma/2$ in the presence of damping (Fig.~1). Moreover, the amplitude of the quantum witness is modulated by the function $f(\varepsilon) = 1-\sqrt{1-\varepsilon^2}$, which follows from the intermediate generalized measurement. For small $\varepsilon$, we have $f(\varepsilon) \simeq \varepsilon^2$ indicating that the witness depends quadratically on the measurement strength for weak measurements. In addition, the derivative $f'(\varepsilon) = \varepsilon/\sqrt{1-\varepsilon^2}$ vanishes for $\varepsilon \rightarrow 0$ and diverges in the opposite limit $\varepsilon \rightarrow 1$ (Fig.~2). The  witness is thus nonanalytic in the strong measurement limit. A small variation of the measurement strength has consequently a  minor influence on the amplitude of the quantum witness \eqref{eq:Wit_weak} for $\varepsilon \rightarrow 0$, whereas it has a significant effect for $\varepsilon \rightarrow 1$. Non-zero values of $\mathcal{W}_q(\tau,\varepsilon)$, that is, nonclassical features, are observed for all values of $\varepsilon>0$ (Fig. 1).

\begin{figure}[t]
\centering
\includegraphics[width=0.45\textwidth]{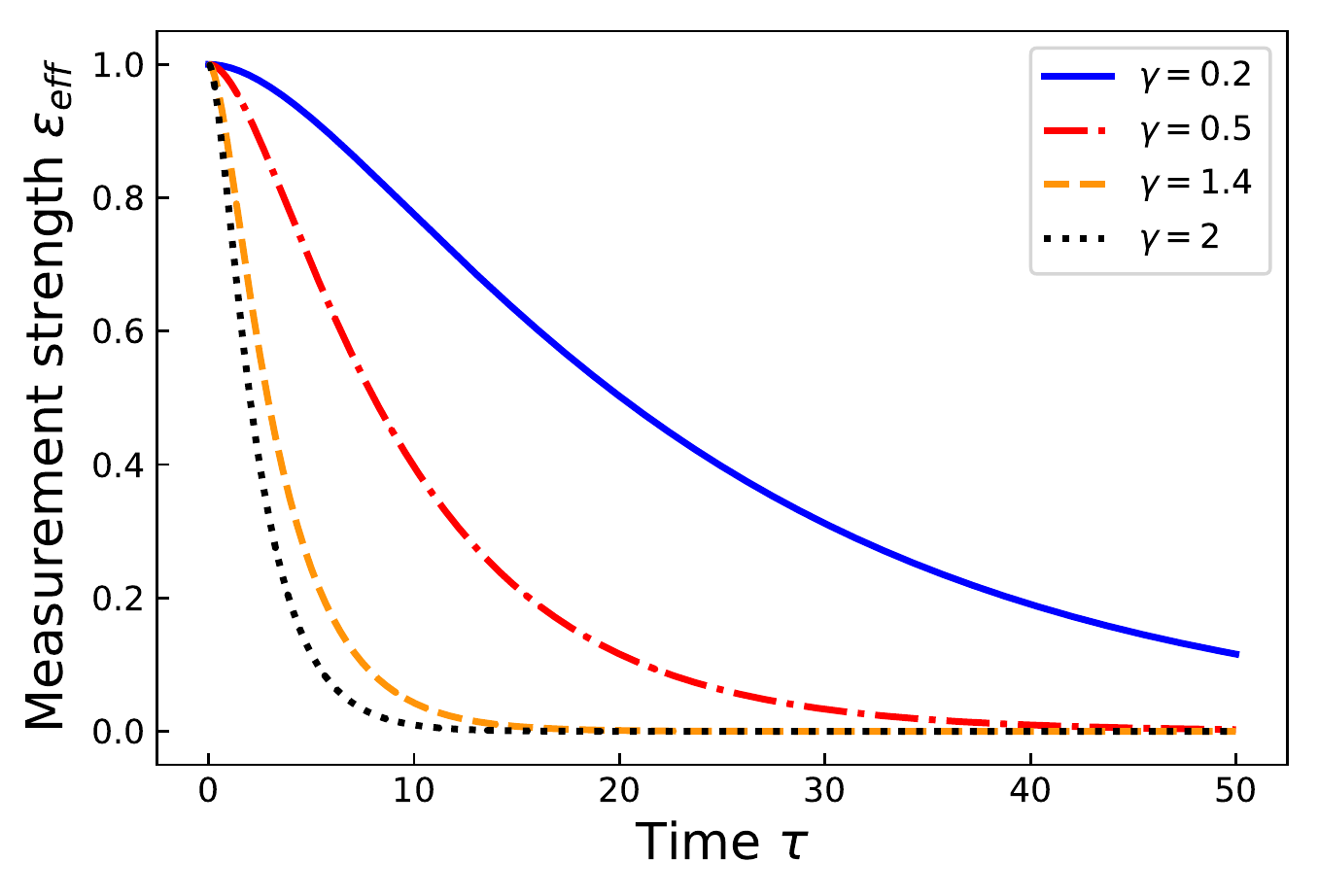}
\caption{Effective measurement strength $\varepsilon_\text{eff}$, Eq.~(14), combining the effect of the  damping and of the finite strength of the generalized measurement, as a function of time $\tau$ for various damping constants. Parameters are $\omega= 1$ and $\varepsilon=1$.}
\label{fig:Weak_Measurement}
\end{figure}

Two factors appear to determine the amplitude of the quantum witness \eqref{eq:Wit_weak}: the external damping and the finite strength of the incomplete measurement. These two effects may be combined by defining an effective measurement strength $\varepsilon_\text{eff}$ so that the amplitude of the quantum witness is given by $f(\varepsilon_\text{eff})/2$. We obtain,
\begin{equation} 
\varepsilon_\mathrm{eff}=\sqrt{1-\left[1-e^{-\frac{\gamma \tau}{2}}\left(1-\sqrt{1-\varepsilon^2}\right)\right]^2}.
\end{equation} In this picture, the coupling to the external reservoir effectively reduces the strength of the measurement to a value that is set by the damping constant and by the time at which the generalized measurement is performed (Fig.~3).

\section{Measurement disturbance}
In order to get a better  understanding of the behavior of the quantum witness described above, it is instructive to relate the amplitude of Eq.~\eqref{eq:Wit_weak} to the so-called measurement disturbance  \cite{bar01,Barnum2002,Maccone2006,mac07}. The latter  quantifies the disturbance, $D=D(\rho,\rho')=1-F(\rho,\rho')$, induced by the measurement device in terms of the fidelity \cite{jos94},
\begin{equation}
\label{14}
 F(\rho,\rho') = \text{Tr}\left[\sqrt{\sqrt{\rho}\rho'\sqrt{\rho}}\right],
 \end{equation}
 between the state $\rho$ before the measurement and the state $\rho'$ after the measurement. The measurement disturbance vanishes, $D=0$, when there is no disturbance, that is, $\rho=\rho'$ and $0< D \leq 1$ when the input state is perturbed by the measurement. We evaluate Eq.~\eqref{14} for a state that    is maximally disturbable by an incomplete measurement described by the operators (7)-(8), for example the $+1$ eigenstate of the operator $\sigma_y$. In that case, the post-measurement state is explicitly given as,
\begin{equation}
\rho'(\varepsilon)=\left(
\begin{array}{cc}
 \frac{1}{2} & -\frac{1}{2} i \sqrt{1-\varepsilon ^2} \\
 \frac{1}{2} i \sqrt{1-\varepsilon ^2} & \frac{1}{2} \\
\end{array}
\right).
\end{equation}
The corresponding pre-measurement state is simply $\rho= \rho'(\varepsilon=0)$. The fidelity between these two states is,
\begin{equation}
F(\rho,\rho')=\frac{1}{2} \left(1+\sqrt{1-\varepsilon ^2}\right).
\end{equation}
As a result, the measurement disturbance is equal to,
\begin{equation}
\label{17}
D=\frac{1}{2} \left(1-\sqrt{1-\varepsilon ^2}\right) = \frac{1}{2}f(\varepsilon).
\end{equation}
Equation \eqref{17} coincides with the amplitude of the quantum witness \eqref{eq:Wit_weak} in the absence of damping. The behavior of the quantum witness as a function of the measurement strength $\varepsilon$ is therefore identical to that of the measurement disturbance of the state that is maximally disturbable by the considered generalized measurement.

\section{Summary}
We have studied the effect  of a generalized perturbing measurement on the quantum witness  based on the no-signaling-in-time condition of a damped qubit. Using the solution of the corresponding quantum master equation, we have explicitly determined the dependence of the quantum witness on the strength of the measurement.  We have found a vanishing derivative  for weak measurements, $\varepsilon \rightarrow0$ and an infinite  derivative for strong measurements, $\varepsilon \rightarrow1$. As a result, the quantum witness is largely insensitive to the measurement strength in the weak measurement regime, while it exhibits a singular, particularly sensitive behavior in the vicinity of projective measurements. This remarkable feature can be directly related to that of   the measurement disturbance defined in terms of the fidelity.

\section{appendix}

We here summarize the derivation of the quantum witness (13) of the damped two-level system. Using the solution (3) of the Lindblad master equation together with Eq.~(4), the density matrix $\rho(\tau)$ of the damped qubit can be written for any given time $t=\tau$ as,
\begin{equation}
\rho(\tau) = \frac{1}{2}\left(\begin{array}{cc} 1-\frac{\gamma_0\left(1-{e}^{-\gamma \tau}\right)}{\gamma} & {e}^{-\frac{\gamma \tau}{2}} e^{-i\omega \tau}\\ {e}^{-\frac{\gamma \tau}{2}} e^{i\omega \tau} & 1+\frac{\gamma_0\left(1-{e}^{-\gamma \tau}\right)}{\gamma} \end{array}\right).
\end{equation}
The probability of finding the system in the state $\left|+\right\rangle$ in the absence of the intermediate nonselective incomplete measurement is accordingly, 
\begin{equation}
P(\left|+\right\rangle) = \left\langle +\right|\rho(\tau)\left|+\right\rangle = \frac{1}{2}\left(1+e^{-\frac{\gamma\tau}{2}}\cos\omega \tau\right).
\end{equation}
The state of the damped qubit after the nonselective generalized measurement at time $\tau/2$ reads,
\begin{equation}
{\rho'}(\tau/2) = A_+\rho(\tau/2)A_+ + A_-\rho(\tau/2)A_-,
\end{equation}
with the  two operators $A_+$ and $A_-$ defined in Eqs.~(9)-(10). We have explicitly,
\begin{widetext}
\begin{eqnarray}
{\rho'}(\tau/2)= \frac{1}{2}\left(\begin{array}{cc} 1-\frac{\gamma_0\sqrt{1-\varepsilon^2}\left(1-e^{-\frac{\gamma\tau}{2}}\right)}{\gamma} & e^{-\frac{\gamma\tau}{4}}\left(\cos\frac{\omega\tau}{2}-i\sin\frac{\omega\tau}{2}\,\sqrt{1-\varepsilon^2}\right)\\ e^{-\frac{\gamma\tau}{4}}\left(\cos\frac{\omega\tau}{2}+i\sin\frac{\omega\tau}{2}\,\sqrt{1-\varepsilon^2}\right) & 1+\frac{\gamma_0\sqrt{1-\varepsilon^2}\left(1-e^{-\frac{\gamma\tau}{2}}\right)}{\gamma} \end{array}\right).
\end{eqnarray}
\end{widetext}
Evolving this state to time $\tau$, the probability to find the system in state  $\left|+\right\rangle$ is given by the expectation value of the projector $\Pi_+$ and is equal to,
\begin{equation}
P'\left(\left|+\right\rangle\right) = \frac{1}{2}\bigg[1 + \,e^{-\frac{\gamma\tau}{2}}\big(\cos^2\frac{\omega\tau}{2}  
- \sqrt{1-\varepsilon^2}\sin^2\frac{\omega\tau}{2}\big)\bigg].
\end{equation}
The quantum witness then follows as,
\begin{equation}
\mathcal{W}_q(\tau,\varepsilon) = 
\frac{e^{-\frac{\gamma\tau}{2}}}{2} \left(1-\sqrt{1-\varepsilon ^2}\right)\sin^2\frac{\omega\tau}{2}.
\end{equation}

\end{document}